\documentclass{tlp}
\usepackage{epsfig, url}

\begin{document}
\bibliographystyle{acmtrans}

\long\def\comment#1{}

\comment{Journal:
  http://journals.cambridge.org/action/displayJournal?jid=TLP and
  latex:
  ftp://ftp.cup.cam.ac.uk/pub/texarchive/journals/latex/tlp-cls/}

\title{ N3Logic: A Logical Framework For the World Wide Web }

\author[T. Berners-Lee]
{ Tim Berners-Lee, Dan Connolly, Lalana Kagal, Yosi Scharf \\
Computer Science and Artificial Intelligence lab\\
Massachusetts Institute of Technology\\
\{timbl, connolly, lkagal, syosi\}@csail.mit.edu
\and Jim Hendler\\
Rensselaer Polytechnic Institute\\
hendler@cs.rpi.edu
}

\submitted{13 May 2006}
\revised{11 Sep 2007}
\accepted{18 October 2007}

\maketitle

\label{firstpage}

\begin{abstract}

The Semantic Web drives towards the use of the Web for interacting
with logically interconnected data. \comment{, geared towards making
  the Web more machine understandable and involves technologies for
  automating the use of Web data. It deals mainly with knowledge
  models that promote common understanding of shared information such
  as Resource Description Framework (RDF).  } Through knowledge models
such as Resource Description Framework (RDF), the Semantic Web
provides a unifying representation of richly structured data.  Adding
logic to the Web implies the use of rules to make inferences, choose
courses of action, and answer questions. This logic must be powerful
enough to describe complex properties of objects but not so powerful
that agents can be tricked by being asked to consider a paradox.  The
Web has several characteristics that can lead to problems when
existing logics are used, in particular, the inconsistencies that
inevitably arise due to the openness of the Web, where anyone can
assert anything. N3Logic is a logic that allows rules to be expressed
in a Web environment. It extends RDF with syntax for nested graphs and
quantified variables and with predicates for implication and accessing
resources on the Web, and functions including cryptographic, string,
math.  The main goal of N3Logic is to be a minimal extension to the
RDF data model such that the same language can be used for logic and
data. In this paper, we describe N3Logic and illustrate through
examples why it is an appropriate logic for the Web.

\comment{ N3 extends RDF with nested graphs, and quantified variables.
  N3 logic adds predicates for implication, look-up on the web and
  built-in functions.  The result is a language which is sound, and
  has the features required and was found to be practically useful.  }

\end{abstract}

\comment{

`` Adding logic to the Web means to use rules to make inferences,
  choose courses of action and answer questions is the task before the
  Semantic Web community at the moment. A mixture of mathematical and
  engineering decisions complicate this task. The logic must be
  powerful enough to describe complex properties of objects but not so
  powerful that agents can be tricked by being asked to consider a
  paradox. Fortunately, a large majority of the information we want to
  express is along the lines of "a hex-head bolt is a type of machine
  bolt," which is readily written in existing languages with a little
  extra vocabulary.  ``
}


\section{Introduction}

\comment{The Semantic Web wants to make data available on the web more
  machine understandable and provide more automation. ..... Provide
  common models for describing meta data about resources and providing
  techniques for retrieving appropriate resources through querying at
  various levels of abstraction. Common models grounded in RDF. More
  automation requires reasoning. Why RDF ? What is the role of URIs ?
}

\comment{The World Wide Web is an open and highly decentralized information
sharing environment.  With no central management (other than the
Internet's Domain Name System for the management of IP addresses),
users are free to share documents without the constraints -- or
benefits -- of peer review, editorial oversight, or even copy editing.
This is both a feature and a problem; it is a feature of a democratic
Web that everyone is free to say what they wish to say (and readers
are free to ignore most of it).  But it is a problem that the
fundamental infrastructure of the Web currently makes no specific
provision to provide the sort of authority metadata on which to base
evaluations of the likely accuracy of information found on the Web.}

The Semantic Web is an enhancement of the World Wide Web to broaden
its sharing capacity to application data sharing as well as
human-focused data sharing. Through RDF the Semantic Web provides a
unifying representation of richly structured data.  And through the
Web, application developers can share these data representations so
that their applications can make ``decisions'' on the basis of combining
the many different kinds of data published on the Web.

RDF builds on the fundamental pointer mechanism of the Web; the
Uniform Resource Identifier (URI).  In the initial incarnation of the
Web, URIs were generally thought to refer to documents and parts of
documents via hypertext anchors. The Semantic Web makes it explicit
that URIs can be used to name anything -- from abstract concepts
(``color'') to physical objects (``the building in which MIT's CSAIL
personnel work'') to electronic objects (``the machine code that
implements the Linux operating system'').  RDF uses URIs to give names
to relationships between objects as well as to the objects themselves.

The abstract representation of RDF \cite{rdfc} is a directed labeled
graph -- i.e. nodes and arcs.  A subset of the nodes are URIs; these
are ``named'' nodes.  Other nodes may not be identifying URIs but the
graph can still describe properties of such nodes including
relationships with other named and un-named nodes.  The properties and
relationships are the edges in the graph.  Every edge type has its own
label and these same edge type labels themselves can be used to name
other nodes in the graph that represent the edge type.  This permits
properties of edge types themselves to be represented in an RDF graph.

N3 is a compact and readable alternative to RDF's XML syntax
\cite{bl01n3}. N3 allows RDF to be expressed but emphasizes
readability and symmetry. It also allows quoting or statements to be
made about statements. This quoting feature allows users to
distinguish between what they believe to be true and what someone else
including a website states or believes. 

N3Logic uses N3 syntax and extends RDF with a vocabulary of
predicates. N3 aims to do for logical information what RDF does for
data: to provide a common data model and a common syntax, so that
extensions of the language are made simply by defining new terms in an
ontology. Declarative programming languages such as Scheme
\cite{scheme} already do this.  However, they differ in their choice
of pairs rather than the RDF binary relational model for data, and
lack the use of universal identifiers as symbols.  N3Logic allows
rules to be integrated smoothly with RDF and provides certain
essential built-in functions that allow information from the Web to be
accessed and reasoned over. The main goal of N3Logic is to make a
minimal extension to the RDF data model so the same language could be
used for logic and data, and to do so in a way that is compatible with
the architectural principles of the Web

In this paper, we discuss the features of N3Logic with the help of
examples and describe its informal semantics.

\section{N3Logic Overview}

One of the main motivations of N3Logic is to be useful as a tool in
the open Web environment. The Web contains many sources of
information, with different characteristics and relationships to any
given reader.  Whereas a closed system may be built based on a single
knowledge base of believed facts, an open Web-based system exists in
an unbounded sea of interconnected information resources. This
requires that an entity be aware of the provenance of information and
responsible for its disposition.  A language for use in this
environment typically requires the ability to express which document
or message asserts what. We found that {\it quoting} provides a
pragmatic solution to this. \comment{The ability to {\it quote}
  logical statement and match them against patterns with variables is
  essential.} However, quotation and reference, with its inevitable
possibility of direct or indirect self-reference, if added directly to
first order logic presents problems such as paradox traps. To avoid
this, N3Logic has deliberately been kept to limited expressive power:
it currently contains no general first order negation.

Another goal of N3Logic is that information, such as but not limited
to rules, which requires greater expressive power than the RDF graph,
should be sharable in the same way as RDF can be shared.  This means
that one person should be able to express knowledge in N3Logic for a
certain purpose, and later independently someone else can reuse that
knowledge for a different unforeseen purpose.  As the context of the
latter use is unknown, this prevents us from making implicit closed
assumptions about the total set of knowledge in the system as a whole.

Further, users of the Web have the ability to express new knowledge
without affecting systems that are already built.  We've chosen to
adopt a monotonicity requirement for N3Logic because we find it scales
well. This implies that the addition of new information from elsewhere
cannot silently change the meaning of the original knowledge, though
it might cause an inconsistency by contradicting the old information.
The non-monotonicity of many existing systems follows from a form of
negation as failure (NAF) in which a sentence is deemed false if it
not held within (or, derivable from) the current knowledge base.  It
is this concept of current knowledge base, which is a variable
quantity, and the ability to {\it indirectly} make reference to it
which causes the non-monotonicity.  In N3Logic, while a current
knowledge base is a fine concept, there is no ability to make
references to it implicitly in the negative.  The negation provided is
called Scoped Negation As Failure (SNAF) and is the ability for a
specific given document (or, essentially, some abstract formula) to
objectively determine whether or not it holds, or allows one to
derive, a given fact.  However, negated forms of many of the built-in
functions are available. (Please refer to Sections \ref{builtin} and
\ref{scoped-negation} for more information about scoped negation and
built-in functions)

\section{Motivating Example}

We describe a Web-based scenario that will be used to illustrate the
different features of N3 logic. Consider a conference management
system that handles different aspects of registration for
conferences. It allows people to register by specifying their names,
addresses, affiliations, and their Friend-of-a-Friend (foaf) page. A
foaf page includes information such as the organization the registrant
works for, her/his current and past projects, and her/his interests
\cite{foaf,foafintro}. Using this information, the conference
management system goes out onto the Web and retrieves relevant
information. By reasoning over this information it is able to make
inferences about the registrant such as whether the registrant is a
vegetarian or not, which workshops she/he would be most interested in,
whether she/he is a member of a certain professional organization, and
whether the registrant is a student. This allows the conference
management system to provide greater support in the registration
process by figuring out what registration fees are applicable, whether
to order vegetarian meals, and by suggesting appropriate workshops.
In the following sections, we define N3Logic in detail, using the
example of this conference system to illustrate key features. 

\section{Notions and Terminology}

N3 is based on the abstract syntax of RDF. The concrete syntax of N3
includes a number of other abbreviations.  Please refer to the
Appendix \ref{appendix} or to the N3 Primer \cite{bl05n3primer} for a
tutorial introduction.

\subsection{Basic Concepts from RDF}

\begin{itemize}

\item{}The atomic formulas in the RDF abstract syntax are called
  ``triples''; they are analogous to one 3-place {\it holds(s, p, o)}
  predicate. For example:

\begin{verbatim}
<http://dig.csail.mit.edu/2006/Papers/TPLP/example/exconf#ExConf>
<http://www.w3.org/1999/02/22-rdf-syntax-ns#type> <http://example.org/conf#Conference>
\end{verbatim}

 \item{}RDF also has conjunctions of formulas. URI terms can be
   abbreviated using namespaces, and the keyword 'a' is short for
   $<$\url{http://www.w3.org/1999/02/22-rdf-syntax-ns#type}$>$.

\begin{verbatim}
  @prefix conf: <http://example.org/conf#> .
  @prefix : <http://dig.csail.mit.edu/2006/Papers/TPLP/example/exconf#> .

  ExConf a conf:Conference .
  ExConf conf:homepage <http://www.l3s.de/~olmedilla/events/MTW06_Workshop.html>.
  ExConf conf:registrant Judy.
  Judy a foaf:Person.
  <http://dig.csail.mit.edu/2005/09/rein/examples/judy-foaf.rdf> foaf:maker Judy.
\end{verbatim}

\item{}RDF also has literal terms; for example:

\begin{verbatim}
ExConf conf:eventName "WWW2006 Workshop on Models of Trust for the Web".
ExConf conf:numOfRegistrants 65.
\end{verbatim}

\item{} Finally, RDF abstract syntax allows existential
  quantification.  Certain syntaxes (e.g. ``[``...'']'') allow an
  existential variable to be introduced without having to name it:
  this is known as a blank node. The following are equivalent:

\begin{verbatim}
  @forSome X. j:Joe foaf:knows X. X foaf:name "Fred" .

  j:Joe foaf:knows [ foaf:name "Fred" ] .
\end{verbatim}

\end{itemize}

\subsection{N3 Extension to RDF}

N3 extends the abstract syntax of RDF in two ways:

\begin{itemize}

\item{}It has all of the terms of RDF plus quoted formulas. For
  example:

\begin{verbatim}
  b:mary says { j:Joe foaf:schoolHomepage <http://example.edu> } .
\end{verbatim}

\item{}It has all of the formulas of RDF plus universally quantified
  formulas. In simple cases, the @forAll quantifier can be left
  implicit. The following are equivalent:

\begin{verbatim}
  @forAll X. { X a Man } log:implies { X a Mortal }.

  { ?X a Man } log:implies { ?X a Mortal }.
\end{verbatim}

\end{itemize}

The N3 example below declares namespace prefixes and defines {\it
  ExConf} an instance of the {\it Conference} class as defined in {\it
  conf} namespace.

\begin{verbatim}
@keywords a.
@prefix conf: <http://example.org/conf#> .
@prefix : <http://dig.csail.mit.edu/2006/Papers/TPLP/example/exconf#> .

ExConf a conf:Conference;
         conf:eventName "WWW2006 Workshop on Models of Trust for the Web";
         conf:acronym "MTW06";
         conf:address "mtw@www.org";
         conf:homepage <http://www.l3s.de/~olmedilla/events/MTW06_Workshop.html> .
\end{verbatim}

\subsection{N3Logic Vocabulary}
\label{builtin}

N3Logic uses the N3 syntax and also includes a set of predicates. Its
vocabulary is union of the N3 syntax and the set of URI references
defined in the log: (\url{http://www.w3.org/2000/10/swap/log#}),
crypto: (\url{http://www.w3.org/2000/10/swap/crypto#}), list:
(\url{http://www.w3.org/2000/10/swap/list#}), math:
(\url{http://www.w3.org/2000/10/swap/math#}), os:
(\url{http://www.w3.org/2000/10/swap/os#}), string:
(\url{http://www.w3.org/2000/10/swap/string#}), and time:
(\url{http://www.w3.org/2000/10/swap/time#}) namespaces as shown in
Table \ref{n3lv}.

\begin{table} [h]
\begin{tabular}{p{12cm}}
\hline log:conclusion, log:content, log:includes, log:semantics,
log:notIncludes, log:supports ... crypto:md5, crypto:sign,
crypto:verify ... list:in, list:last ... math:lessThan,
math:greaterThan ... os:argv, os:environ ... string:contains,
string:endsWith, string:scrape ... time:day, time:hour, time:minute
...\\

\hline
\end{tabular}
\label{n3lv}
\caption{Some N3Logic predicates}
\end{table}

While N3Logic properties can be used simply as ground facts in a
database, is very useful to take advantage of the fact that in fact
they can be calculated. N3Logic includes axiom schemas for each of
these terms; reasoners can use these axioms to evaluate formulas and
bind variables.  These are called {\it built-in functions} and they
can be used to provide a variety of functionality, for example the
crypto:sha1 built-in allows the object to be computed as the SHA-1
hash of the subject.

\section{Informal Semantics of N3Logic}

Various vocabularies, notably RDFS and OWL, have defined RDF
predicates with logical semantics, such as rdfs:range, and owl:sameAs,
etc.  Of these, N3 logic uses rdf:type and owl:sameAs, and defines
further predicates to allow rules, web access and built-in calculated
functions. N3Logic extends RDF in two ways (i) a syntax namely N3, and
(2) a vocabulary of new predicates, which can be used to talk about
the provenance of information, contents of documents on the Web, and
provide a variety of useful functionality such as string,
cryptographic, and mathematic functions.

\comment{Just as OWL \cite{owl} is expressed in RDF by defining
  properties, in the same way queries, differences, and such functions
  can be expressed in N3 with N3Logic.}

N3Logic allows statements to be made about, and to query, other
statements such as the content of data in information resources on the
Web. Formulae provide the ability to represent such sets of
statements.  To allow statements about them, some of the relationships
defined are given URIs so that these statements and queries can be
written in N3.

The fact that the rule language and the data language are the same
gives a certain simplicity (there is only one syntax) and completeness
(rules can operate on themselves, anything written in N3 can be
queried in N3). This would be broken if a special syntax were added
for built-in functions and operators. Instead, these are simply
represented as RDF properties.  Rules may have full N3, even with
nested graphs, on both sides of the implication. This gives a form of
completeness as rules can generate rules.

\subsection{Relationship with RDF, RDFS, and OWL}
\label{relation}

N3 syntax allows RDF to be expressed, however, it does not make use of
the full RDF vocabulary. 

\begin{itemize}

\item{} In N3 the keyword ``a'' is a shorthand for rdf:type and can be
  replaced with a direct use of the full URI symbol for {\it
    rdf:type}.

\comment{\item{}The predicate in a triple, \{ s p o \}, is
  automatically inferred to be of type {\it rdf:Property} even if it
  is not explicitly declared as one.

\item{} The object of a triple defined as \{ s rdf:type o \} or \{ s a
  o \} where ``a'' is the keyword for {\it rdf:type } is inferred to be
  of type {\it rdfs:Class}. }

\item{} The shorthand notation of ``='' refers to {\it owl:sameAs}. It
  is used to state that the subject and object are equal.

\item{}N3 uses {\it rdf:List}, {\it rdf:first}, {\it rdf:rest}, and
  {\it rdf:nil} for describing lists. Implementations may treat list
  as a data type rather than just a ladder of rdf:first and rdf:rest
  properties. The use of rdf:first and rdf:rest can be seen as a
  reification of the list datatype.  This use of lists requires more
  axioms than are actually defined in the RDF specification. These
  axioms, described in N3, are given below 

\begin{itemize}

\item{Existence of lists}
\begin{verbatim}
 (?X).
 { ?L rdf:rest [ ] } => { [ ] rdf:first ?X; rdf:rest ?L }.
 \end{verbatim}

\item{Uniqueness of lists}
\begin{verbatim}
{ ?L1 rdf:first ?X; rdf:rest ?R.
  ?L1 rdf:first ?X; rdf:rest ?R. } => { ?L1 = ?L2 }.
\end{verbatim}

\item{Uniqueness of the rdf:first and rdf:rest of a list}
\begin{verbatim}
{ ?S rdf:first ?O1, ?O2 } => { ?O1 = ?O2 }.
{ ?S rdf:rest ?O1, ?O2 } => { ?O1 = ?O2 }.
\end{verbatim}

\end{itemize}

\item{} The semantics of RDFS can be easily expressed in N3Logic. A
  set of N3Logic rules for defining {\it rdf:domain} and {\it
    rdf:range} are as follows

\begin{verbatim}
{ ?S [ s:domain ?C ] ?O } => { ?S a ?C } .
{ ?S [ s:range ?C ] ?O } => { ?O a ?C } .
{ ?S a [ s:subClassOf ?C ] } => { ?S a ?C } .
\end{verbatim}

\end{itemize}

\subsection{Quoted N3 Formulae}

\comment{An N3 formula is any valid N3 expression and may be comprised
  of a conjunctive set of triples and formula literals (quoted
  formulae).}

Quoting is an important feature provided by N3.  Various forms of
literal value are allowed in RDF graphs, however the RDF standard does
not itself provide for another RDF graph to be a data value. Remedying
this allows one to express relationships between graphs, for example
that a given graph is the RDF content of a particular document. Every
RDF graph is a subclass of N3 formula. A quoted N3 formula is an N3
formula enclosed in braces ``\{`` ``\}''. Some examples of N3 formulae and
quoted N3 formulae are shown in Table \ref{n3formula}.

\comment{Substitution of variables is defined to recursively apply
  inside formulae, as is usual. However, substitution does not occur
  recursively with respect to the {\it owl:sameAs} property.}

Note that, however, substitution of equal (owl:sameAs) terms occurring
in a formula does not take place within nested quoted formulae:
formulae are not referentially transparent.

In our conference management example, we assume that there are several
administrators. Each administrator specifies a list of people who may
not register due to some reason. The example below states that Joe
says that Peter is not permitted to register. Even though Peter is
equal to John, it does not imply that Joe says that John is not
permitted to register. This is an example of a quoted N3 formula.

\begin{verbatim}
j:Joe says { mit:Peter policy:notpermitted conf:Register }.
mit:Peter = cmu:John.
\end{verbatim}

\begin{table} [!th]
\begin{tabular}{p{12cm}}

\hline 

N3 formula : c1 is a Car as defined in the ``ex'' namespace and its
color is green.

\begin{verbatim}
@keywords a.
@prefix ex:  <http://example.org/car.n3#> .
c1 a ex:Car;
      ex:color  "green".\end{verbatim}\\

\hline 

N3 formula : The semantics of
$<$\url{http://www.example.org/myfoaf.rdf}$>$ as expressed in N3 is
stored into a variable, F.

\begin{verbatim}
@forAll F.
<http://www.example.org/myfoaf.rdf> log:semantics F . 
\end{verbatim}\\

\hline

Quoted N3 formula : Joe believes that Peter is a graduate student.

\begin{verbatim}
j:Joe believes { mit:Peter a school:GraduateStudent } .
\end{verbatim}\\

\hline

Quoted N3 Formula : Mary believes that Joe believes that Peter is a
graduate student.

\begin{verbatim}
Mary believes { j:Joe believes { mit:Peter a school:GraduateStudent } } .
\end{verbatim}\\

\hline
\end{tabular}
\caption{N3 Formula Examples}
\label{n3formula}
\end{table}

\subsection{Dereferencing URIs}

The Web is exposed as a mapping between URIs and the information
returned when such a URI is dereferenced, using appropriate protocols.
In N3Logic, the information resource is identified by a symbol, which
is in fact is its URI. The information is represented in formulae, the
information retrieved is also represented as a formula.  Not all
information on the Web is, of course in N3. If we assume that N3 is
the interlingua, then from the point of view of this system, the
semantics of a document is exactly what can be expressed in N3.

{\it log:semantics} is a logical property that represents the relation
between a document and the logical expression which represents its
meaning expressed as N3. The {\it log:semantics} of an N3 document is
the formula achieved by parsing the representation of the document.

In order to access a foaf file and store its N3 representation to a
variable, {\it log:semantics} is used in the following manner

\begin{verbatim}
<http://www.example.org/myfoaf.rdf> log:semantics ?F. 
\end{verbatim}

The Architecture of the World Wide Web defines algorithms by which a
machine can determine representations of document given its symbol -
URI.  For a representation in N3, this is the formula which
corresponds to the document production of the grammar.  For a
representation in RDF/XML, it is the formula which is the entire graph
parsed.  For any other languages, it may be calculated as long as a
specification exists that defines the equivalent N3 semantics for
files in that language. The N3 semantics of other languages for Web
documents such as GRDDL \cite{grddl} and RDF/A \cite{rdfa} may be
defined so that they are also calculable.

The N3 formula of a document is not the semantics of the document in
any absolute sense.  It is the semantics expressed in N3.  In turn,
the full semantics of an N3 formula are grounded in the definitions of
the properties and classes used by the formula.  In the HTTP space, in
which URIs are minted by an authority, definitive information about
those definitions may be found by dereferencing the URIs. This
information may be in natural language, in some machine-processable
logic, or a mixture.  \comment{Two patterns are important for the
  Semantic Web.  One is the grounding of properties and classes by
  defining them in natural language.  Natural language, of course, is
  not capable of giving an absolute meaning to anything in theory, but
  in practice a document carefully written by a group of people
  achieves a precision of definition that is quite sufficient for the
  community to be able to exchange data using the terms concerned.
  The other pattern is the raft-like definition of terms in terms of
  related neighboring ontologies. }

\begin{figure}[tb]
  \centerline{\epsfig{file=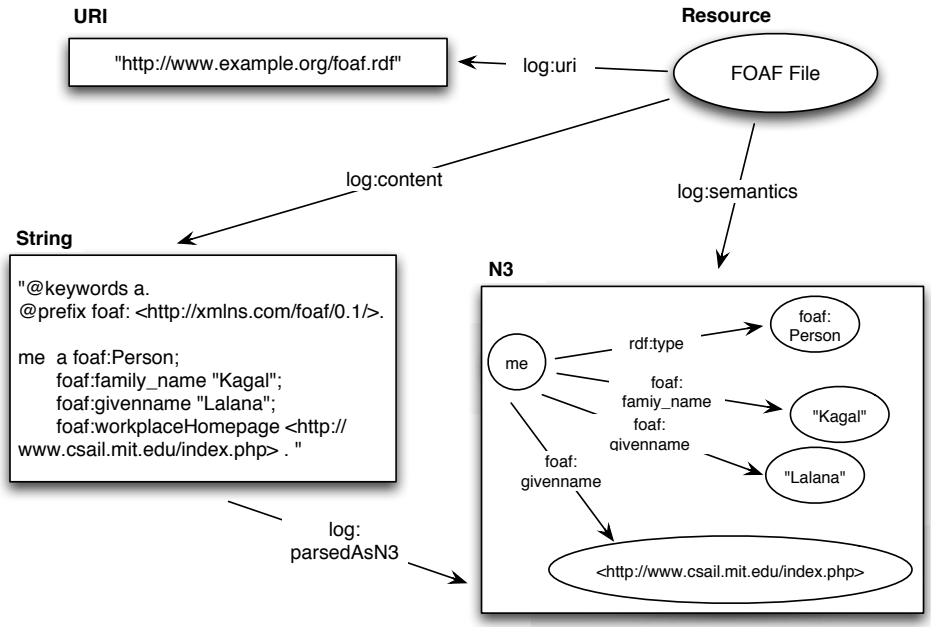, height=3.5in}}
  \caption{Relationship between resource, URI, string, N3
    representation, and logic properties.}
  \label{relationship}
\end{figure}

Other logical properties related to dereferencing include {\it
  log:content}, {\it log:parsedAsN3}, {\it log:N3String}, and {\it
  log:uri}.

\begin{itemize}

\item {\it log:content} of a document, is the string which was
  returned as the document contents when the URI of the resource was
  looked up.

\item {\it log:parsedAsN3} is the logical property that returns the
  formula got by parsing the string as a Notation3 document.

\item {\it log:N3String} returns the string which expresses a certain
  formula in N3.

\item {\it log:uri} is the URI of a resource. Normal logic processing
  should not look at URIs but in some cases one needs to. Though a URI
  itself has semantics and can be used to obtain additional
  information, log:uri must be used carefully because it is a
  level-breaker: it lets an N3 system look at its infrastructure. It
  is a function which one can evaluate either way: resource to URI or
  URI to resource.

\begin{verbatim}
<http://example.org/> log:uri "http://example.org/"
\end{verbatim}

\end{itemize}

Figure \ref{relationship} illustrates the relationship between URI,
resource, string, N3 representation, and the different logical
properties.

\comment{
\subsection{Variables} 

N3Logic allows both existential and universal variables to be declared
and used in rules. These quantifiers have the usual qualities.
Existential quantifiers indicates that, where the formula is
considered true, it is true for at least one substitution mapping of
the existential variables onto non-variables whereas universal
quantifiers indicate that, it is true for every substitution mapping.
}

\comment{Within a formula F, ?x is short for :x, an existential
  variable and does not need to be explicitly declared. But implies
  @forAll :x on the formula within which F is nested.}

\subsection{Rules} 
\label{rules}

N3 extends RDF with variables and nested graphs to enable the
descriptions of rules. In its blank nodes (items in the graph not
directly identified by a URI) an RDF graph has a form of existential
variable. Extending the language to allow variables existentially or
universally quantified over a graph allows N3 to be used for a form of
logic. The drive for this initially was so that, given variables, a
rule is just a relation between two graphs. Variables are defined such
that when substitution occurs in a graph, it also occurs in any nested
graph. 

The {\it log:implies} property expresses a rule, its subject being the
antecedent graph, and the object being the consequent graph. It
relates two formulae, expressing implication. A statement using {\it
  log:implies} cannot be calculated.  It is not a built-in function,
but a predicate which allows the expression of a rule.  N3 allows
blank nodes in the conclusion of a rule, hence allowing the creation
of new objects.

Continuing with our earlier example, though anyone can specify a list
of people who cannot register, the conference management system only
trusts administrators of the system with this information.

\begin{verbatim}
@forAll A, X. 
{ A a conf:Administrator.
  A conf:says { X policy:notpermitted conf:Register }.
} => { X policy:notpermitted conf:Register }.
\end{verbatim}

To do some inference within another set of rules, a useful
relationship is {\it log:conclusion}. It is a property between a
formula, and the result of running any rules in the formula on all the
data recursively - the deductive closure. To make up the initial
formula, {\it log:conjunction} can be used to merge a list of
formulae.

For example, if the statements made by administrators is in file
$<$statements.n3$>$, the list of administrators in file
$<$admin.rdf$>$, and the rule described about in $<$rule.n3$>$, then
we can use {\it log:conjunction} and {\it log:conclusion} to decide
whether or not someone should be permitted to register.  {\it
  log:supports} is a relationship that combines {\it log:conclusion}
and {\it log:includes} to check whether the conclusion of the formula
includes a certain subgraph. (Please refer to section
\ref{scoped-negation} for more information about {\it log:includes}).

\begin{verbatim}
@forAll S, A, R, C, X.
{ <statements.n3> log:semantics S.
  <admin.rdf> log:semantics A.
  <rule.n3> log:semantics R.
  (S A R) log:conjunction C.
  C log:supports { X policy:notpermitted conf:Register }.
} => { X policy:notpermitted conf:Register }.
\end{verbatim}

As it is possible to have blank nodes in the conclusion of a rule,
{\it log:conclusion} and {\it log:supports} are undecidable and may
run forever. However, it is possible to restrict N3Logic to a
decidable subset language in which blank nodes are not allowed in the
conclusion.

\comment{

Tim:
>Why semi-decidable? Isn't log:includes decidable graph-isomorphism
>complete?  (modulo builtins which have their own decidability)  It
>is just graph matching.
>

Jim :
I thought that log:includes included what was entailed from the
document, not just the text inclusion, but maybe that was
log:semantics  - I thought you and I discussed this when I visited
you and decided that it could be (cwm is a forward-chainer and I can
set up a rule set that will keep it going forever -- either way, we
need to say something about this

Tim :
Ah.  No, log:includes is ~ textual, "N3 entailment".

The things which involve running the rule engine are log:conclusion  
(deductive closure) and log:supports (derivability, implemented in  
cwm but not necessarily in Euler as a combination
of log:conclusion and log:includes.)

Yes, because of bnodes in the conclusion of a rule, cwm can run
forever in these cases.  There is a subset language in which bnodes
are not allowed in the conclusion.  (datalog N3 and horn n3?)

}

\subsection{Graph Functions}
\label{scoped-negation}

The logic function, {\it log:includes}, checks whether one formula can
be {\it N3-derived} from another formula. (Please refer to Section
\ref{n3-derivation} for more information about N3-derivation)
Together, {\it log:semantics} and {\it log:includes} allow rules to
access the Web, and to objectively check the contents of documents,
without having to load them and believe everything they say.

The following rule states that if the home page of a registrant says
that she is a vegetarian, then she/he is a vegetarian. We find the URI
of her homepage on her foaf page.

\begin{verbatim}
@forAll X, FOAF, F, H, HS.
{ ExConf conf:registrant X.
  FOAF foaf:maker X.
  FOAF log:semantics F.
  F log:includes { X foaf:homepage H }.
  H log:semantics HS.
  HS log:includes { X a ex:Vegetarian }
} => { X a ex:Vegetarian}.
\end{verbatim}

Whereas some datasets (such as a list of members of a club) are
definitively complete, others (such as a set of temperature
measurements) are not. This aspect of the Semantic Web makes negation
as failure meaningless unless it is associated to a specific dataset.
As a formula is of finite size, it can be tested for what it does not
say. As a form of negation, {\it log:notIncludes} is completely
monotonic.  It can be evaluated by a mathematical calculation on the
value of the two terms: no other knowledge gained can influence the
result.  This is the {\it scoped negation as failure} mentioned in the
introduction.  This is not a non-monotonic negation as failure.

\begin{figure}[tb]
  \centerline{\epsfig{file=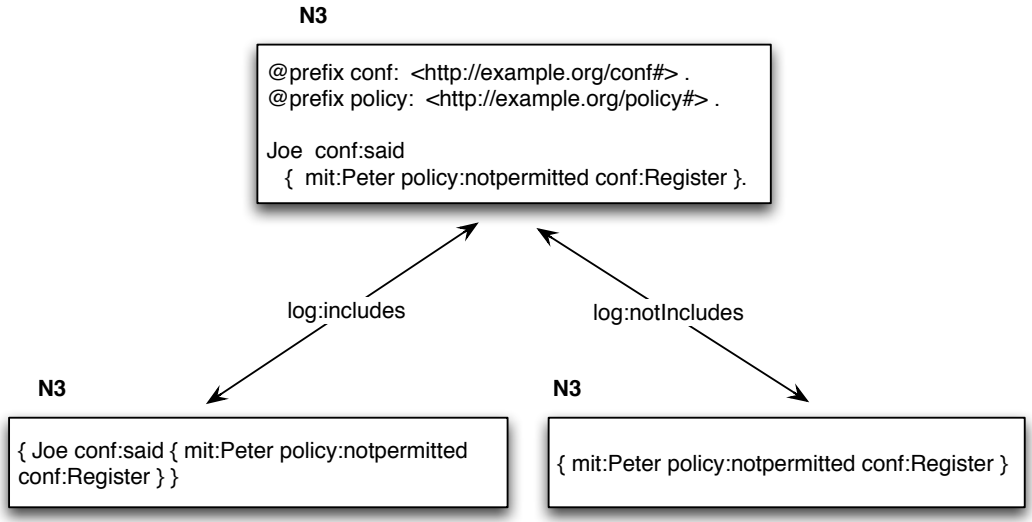, height=3in}}
  \caption{Quoting and its relation to log:includes and
    log:notIncludes}
  \label{includes-not}
\end{figure}

Figure \ref{includes-not} shows how {\it log:includes} and {\it
  log:notIncludes} relate to quoting. 

We continue with our example. We assume that every school website has
a property linking to its directory. The directory provides
information about people including their foaf pages, their designation
(such as student, faculty, associate), and their email address. Below
we have a rule that states if the directory of the school does not
specify that the person under consideration is a student, the system
gives the person the academic rate but not the student rate.

\begin{verbatim}
@forAll X, FOAF, F, H, HS, D, DS.
{ ExConf conf:registrant X.
  FOAF foaf:maker X.
  FOAF log:semantics F.
  F log:includes { X foaf:schoolHomepage H }.
  H log:semantics HS. 
  HS log:includes { H school:directory D }.
  D log:semantics DS.
  DS log:notIncludes { X a school:Student }
} => { X conf:registrationRate conf:AcademicRate }.
\end{verbatim}

The effect of a default with an explicit domain is achieved with {\it
  log:notIncludes}. Defaults can be handled by first running rules to
work out everything that is specified, and then doing a {\it
  log:notIncludes} on the result as shown in the example above.

\subsection{Built-ins}

N3Logic also provides other built-ins for additional
functionality. Some examples include

\begin{itemize}
\item{} {\it crypto} functions - md5, sign, and verify
\item{} {\it xmath} functions - cos, greaterThan, notGreaterThan, and
  difference
\item{} {\it os} functions for retrieving environment information -
  argv, baseAbsolute, and baseRelative
\item{} {\it string} functions - concatenation, matches, and
  startsWith
\item{} {\it time} functions - dayOfWeek, gmTime, and localTime
\end{itemize}

The following example describes a rule that states that papers
submitted to the conference that are not more than 6 pages are valid
if authorized by a program chair of the conference.

\begin{verbatim}
@prefix math: <http://www.w3.org/2000/10/swap/math#> .

@forAll PAPER, LEN, CHAIR.
{ ExConf conf:submittedPaper PAPER.
  PAPER conf:pageLength LEN.
  LEN math:notGreaterThan 6.
  PAPER conf:authorized CHAIR.
  CHAIR conf:chair ExConf.
} => { PAPER a conf:ValidPaper }.
\end{verbatim}

\comment{The following example describes a rule that states that checks that
are not more than \$10,000 are valid if signed by a manager from the
Finance department.

\begin{verbatim}
@forAll CHECK, AMT, MANAGER.
{  CHECK a biz:Check.
   CHECK biz:amount AMT.
   AMT math:notGreaterThan 10000.
   CHECK biz:signedBy MANAGER.
   MANAGER a biz:Manager; biz:department biz:FinanceDept.
} => { CHECK a biz:ValidCheck }
\end{verbatim}
}

\subsection{N3 Derivation}
\label{n3-derivation}

{\it N3-derivation} is not aimed at moving all the way to the powerful
inference capabilities of an expressive logic, but rather extends
textual inclusion to include some simple inferences that are standard
in most logics.  These include:

\begin{itemize}

\item{Conjunction Elimination (CE): } For any formulae A and B, given
  conjunction A and B, then A follows, and B follows.

\item{Conjunction Introduction (CI): } For any formulae A and B, given
  A, and given B, then the conjunction A and B follows.

\item{Universal Elimination (UE): } Given any formula A, a universal
  variable x that is used in A, and a ground term t, A{$_t/x$}
  follows. i.e. @forAll x, A and ground term t, then A{$_t/x$}
  follows.

\item{Existential Introduction (EI): } Given any formula A containing
  a ground term t, and an existential variable v that does not occur
  in A, then A{$_x/t$} follows.

\item{Variable Renaming (VR): } For any formula A, and variables x and x',
  A = A' where A' is a formula derived by Subst({x/x'}, A).

\end{itemize}

N3 derivation is any finite number of applications of of CE, CI, UE,
EI or VR. An N3 derivation operator {\it T} is defined as any operator
which is the successive application of any sequence (possibly empty)
of such operators.  A formula F n3-derives a formula {\it T} F, which
implies that by a combination of CE, EI, UE and VR, {\it T} F
logically follows from F.

As RDF Graph is a subclass of N3 formula, if F and G are RDF graphs,
only CE and EI apply and n3-derivation reduces to simple entailment
from RDF Semantics.

The implementation of built-in functions is not in general required
for any implementation of N3Logic, as they can always soundly be
treated as ground facts.  However, their usefulness is derived from
their implementation. For example, \{ 1 math:negation -1 \} is derived
by calculation.  Like other RDF properties, the set is designed to be
extensible, enabling others to use URIs for new functions.  When a
triple can be evaluated, or a variable bound because its predicate is
a built-in function, then the derivation of the statement is said to
be by {\it calculated derivation}. {\it N3Logic-derivation} is
N3-derivation with modus ponens and calculated derivation.

\comment{
\begin{itemize}
\item{Modus Ponens : } For any formulae, A and B, if A =$>$ B, and A is
  true, then B is true.
\end{itemize}
}

\subsection{Symmetry of Triples}

When designing an ontology in RDF the direction chosen for a given
property is arbitrary - one can either define ``parent'' or ``child'',
``employee'' or ``employer''. Our philosophy (from the Enquire design
of 1980 \cite{enquire}) is that one should not favor one way over
another.  On the other hand, one should not encourage people to
declare both a property and its inverse. Therefore, a design choice in
N3 is to allow {\it forward and backward links to be expressed with
  equal ease}. It does this by providing keywords ``is'' and ``of''
that allow one to reverse the direction of the description of a
triple. This also enables the serialization of any acyclic graphs with
blank nodes without requiring them to have generated node identifiers.

\subsection{Extensibility}

The extensibility of RDF is deliberate so that a document may draw on
predicates from many sources.  The statement \{s p o\} expresses that
the relationship denoted by p holds between the things denoted by s
and o.  The meaning of the statement \{s p o\} in general is defined
by any specification for p. The Architecture of the Web specifies
informally how information about the relation can be discovered
\cite{awww}. In a similar fashion, N3Logic allows external predicates
- predicates not defined within N3Logic - to be used. The definitions
of these external predicates can be discovered by looking up their URI
on the Web and used as long as their semantics are defined in
N3Logic. Clearly a system which includes further logical predicates,
beyond those defined in N3Logic, whose meaning introduces greater
logical expressiveness would change the properties of the logic.

By having rules and data in the same languages, N3Logic provides
simplicity in syntax and completeness as rules can operate on
themselves and anything written in N3 can be queried in N3. A rule
engine, when analyzing a rule prior to running it, can treat specially
those properties it knows as calculable functions which occur in the
antecedent. This allows N3 to be used to develop specific languages
such as query languages. For example, we can create a language for
expressing graph differences and updates by simply defining two new
properties, {\it diff:insertion} and {\it diff:deletion}
\cite{bl04delta}. These properties provide a way to uniquely identify
what is changing and a way to distinguish between the pieces added and
those subtracted.

In the following example, everyone who is paying a student rate, will
also be given accommodation in the dormitory but will not be given a
ticket to attend the social event of the conference.

\begin{verbatim}
@forAll X.
{ X conf:registrationRate conf:StudentRate } 
diff:insertion { X conf:accommodation conf:Dormitory };
diff:deletion { X conf:ticket conf:SocialEvent } .
\end{verbatim}

In many languages similar to N3, there is a risk of ambiguity as to
whether a naked alphanumeric string is a keyword or an
identifier. Serious version management problems occur when new
keywords are added to a language, changing things which were
identifiers into keywords. N3 is designed to be extended in the
future. For this reason, an N3 document can declare which keywords it
uses without the ``@'' sign. This allows N3 to be extended without the
danger that existing documents be incorrectly interpreted by future
systems, or future documents by existing systems.

\section{N3Logic Example: Access control policy}
We extend our conference management system example with a policy for
controlling access to pictures that were taken during the
conference. This policy states that {\it only people who registered
  for the conference can view pictures taken at the conference}.
During the registration process, the system records registrant's foaf
pages. In order to prove that they have registered for the conference,
users must be able to prove ownership of a registered foaf page. This
can be done either using a decentralized authentication mechanism such
as Open ID \cite{openid} or using a mechanism by which users must
present a secret key whose hash is on their foaf page. In this policy,
we use the latter. A user request consists of her secret key and the
URI of the picture being requested. If the picture is one of the
pictures taken at the conference and the secret key is the digest of
the {\it session:hexdigest} value on a registered foaf page, then the
request is considered valid and the picture is returned to the user.
For the entire working example, please refer to
\url{http://dig.csail.mit.edu/2006/Papers/TPLP/example/}.

\begin{verbatim}
@forAll REQ, PHOTO, WHO, FOAF, X, TXT, CONF, C.

{ REQ a rein:Request.
  REQ rein:resource PHOTO.
 <http://dig.csail.mit.edu/2006/Papers/TPLP/example/exconf.n3> log:semantics C.
  C log:includes
        { CONF a conf:Conference. PHOTO a conf:GroupPicture; conf:taken CONF }.

  REQ rein:requester WHO.
  WHO session:secret ?S.
  ?S crypto:md5 TXT.

  C log:includes
        { CONF conf:registrant X. FOAF foaf:maker X }.
  FOAF log:semantics [ log:includes
        { FOAF foaf:maker [ session:hexdigest TXT ] }
    ].

} => { REQ rein:requester [ policy:permitted-to-view PHOTO ]. REQ a ValidRequest }.
\end{verbatim}

\section{Implementations}

We have developed cwm \cite{bl00cwm} a forward-chained
reasoner in python \cite{python} for N3 and N3Logic. It is a
general-purpose reasoner for the Semantic Web that can be used for
querying, checking, transforming and filtering information. Currently,
cwm parses RDF/XML, and N3 and its subsets. A number of tools, for
example SWOOP \cite{kalyanpur05swoop}, support Turtle
\cite{beckett03turtle}, a fragment of N3 that is equivalent to
RDF/XML.

Being based on a more expressive logic language adds a host of
features to cwm not available to other RDF processing tools: accessing
Web resources, and filtering RDF graphs after merging them, for
example. Since N3Logic is expressive enough so that positive
datalog-like rules can be expressed in it, cwm is able to reason using
a first order logic but without classical negation. Combining this
reasoning functionality with its ability to retrieve documents form
the Web as needed, the system can be considered a reasoner for the
Web. It has grown from a proof of concept application to a popular
rule engine, used in major research projects such as Policy Aware Web
\cite{kolovski05paw,kagal06aaai}, and the Technical Report Automation
project at W3C (\url{http://www.w3.org/TR/}).

Euler is an inference engine supporting logic based proofs. Unlike
cwm, it is a backward-chaining reasoner enhanced with Euler path
detection \cite{euler}.

Pychinko is a Python implementation of the classic Rete pattern
matching algorithm \cite{pychinko}. Rete has shown to be, in many
cases, the most efficient way to apply rules to a set of facts--the
basic functionality of an expert system. Pychinko employs an optimized
implementation of the algorithm to handle facts, expressed as triples,
and process them using a set of N3 rules. Pychinko tries to closely
mimic the features available in Cwm, as it is one of the most widely
used rule engines in the RDF community. Pychinko has proven to be
faster than Cwm, however it's limitation lies in its expressivity:
Pychinko cannot handle most of the cwm built-ins. It is worth
mentioning here that the RETE engine used in Pychinko has been ported
to Cwm - thus Cwm can now boast the same performance improvements.

\comment{FuXi is another implementation of N3Logic \cite{fuxi}. The
  authors were mainly interested in rules that could represent various
  Description Logic relationships. FuXi is a Rete engine built on top
  of rdflib \cite{rdflib}.  It supports no builtins, but, as part of
  python-dlp[1], has code to generate rules to process the OWL that
  appears in a set of documents.}

\section{Related Work}

Several logics related to N3Logic exist including OWL, Simple Common
Logic (SCL) \cite{scl}, and Knowledge Interface Format (KIF)
\cite{kif}. OWL is built on top of RDFS and provides a vocabulary for
describing the characteristics of properties and classes, the
relationships between classes, and relationships between
properties. OWL is based on Description Logic \cite{dl}, which is a
subset of First Order Logic (FOL) \cite{shapiro05logic}.  OWL provides
limited expressivity for a Web-like environment as it does not support
quantified variables, rules, or a mechanism to distinguish which
document or person asserts what. KIF is a framework for exchanging of
declarative knowledge among heterogeneous computer systems. It is a
version of first order predicate calculus with extensions to support
non-monotonic reasoning and quoting. The key differences between KIF
and N3Logic are that KIF does not include operators for Web access and
it supports non-monotonic reasoning. SCL is aimed at providing a
standard logical interchange language based on XML. It has a
higher-order syntax that provides integration between different
representation languages but SCL gives this syntax a completely
first-order semantics. This syntax can be used to provide quoting
functionality. Proof Carrying Authorization (PCA) proposes that the
underlying framework of a distributed authorization system be a
higher-order logic and that different domains in this system use
different application-specific logics that are subsets of the
higher-order logic \cite{appel99pca}. They also propose that clients
develop proofs of access using these application specific logics and
send them to servers to validate. N3Logic draws inspiration from PCA
but modifies it to leverage the distributed nature and linkability of
the Web.

A formal categorization of N3Logic is complicated as it differs from
most traditional logics in expressivity.  It is clearly more
expressive than Datalog \cite{gallaire78logic} but less expressive
than traditional FOL.  Much Semantic Web work uses DL expressivity,
and like DL, N3Logic is a subset of FOL, although the quoting
mechanisms provide higher order features (which we believe are
actually limited to FOL in the same manner as SCL).  However, unlike
DL, N3Logic is not decidable, limiting expressivity in other ways
motivated by the Web considerations discuss in this paper.  As such,
developing a formal model theory for N3Logic is quite challenging, and
is the focus of current work.

\section{Conclusion}

The main goal of N3Logic is to extend the RDF data model, so that the
same language can be used for logic and data. N3Logic uses the N3
syntax, which provides quoting, variables, and the implication
operator. N3Logic also includes built-in functions that allow rules to
access Web resources, define which inference can be drawn from
specific Web documents, and other useful functionality such as
mathematic, cryptographic, and string. In this paper, we described the
N3 syntax and give the informal semantics of N3Logic.

The use of {\it log:notIncludes} to allow default reasoning without
non-monotonic behavior achieves a design goal for distributed rule
systems. The N3Logic language has been found to have some useful
practical properties.  The separation between the N3 extensions to RDF
and the logic properties has allowed N3Logic to be extended with other
properties to provide functionality such as the expression of graph
differences and updates \cite{bl04delta}.

\section{Acknowledgements}
We are grateful to Vladimir Kolovski, Sean Palmer, Dave Beckett, and
Jos de Roo for feedback on the N3 language resulting from their
implementations; to the Data Access Working Group for feedback
resulting from their adoption of N3 syntax for part of SPARQL grammar;
to the RDF working group for their co-operation in keeping NTriples a
subset of N3; and to many in the W3C Semantic Web Interest Group for
helpful advice and suggestions. This work is supported in part by
funding from US Defense Advanced Research Projects Agency (DARPA) and
Air Force Research Laboratory, Air Force Materiel Command, USAF, under
agreement number F30602-00-2-0593, Semantic Web Development. This work
was also funded under NSF ITR 04-012.

\bibliography{n3logic-tplp}

\section{APPENDIX : N3 Concrete Syntax}
\label{appendix}

N3 provides a human-readable syntax for RDF and is a language that
uses conventional unix-style punctuation, which is both more easily
writable and readable than the RDF/XML syntax \cite{bl06n3bnf}. It
provides quantified variables and allows quoting so that statements
about statements can be made.

\begin{itemize}

\item{} It provides URI abbreviation using prefixes which are bound to
  a namespace. 

\begin{verbatim}
@prefix j: <http://example.org/joe-foaf#> .
@prefix rdf: <http://www.w3.org/1999/02/22-rdf-syntax-ns#> .
@prefix rdfs: <http://www.w3.org/2000/01/rdf-schema#> .
@prefix conf: <http://example.org/conf#> .
@prefix : <http://dig.csail.mit.edu/2006/Papers/TPLP/example/exconf#> .
\end{verbatim}

The prefixes used in the above example are assumed throughout the
paper.

\item{} Qualified names using the default namespace have an empty
  prefix and start with a colon.
\begin{verbatim}
@prefix : <http://dig.csail.mit.edu/2006/Papers/TPLP/example/exconf#> .

ExConf a conf:Conference .
\end{verbatim}

\item{} An N3 formula is comprised of a conjunctive set of triples.

\item{} Triples are formed of terms, which are symbols (URIs),
  strings, blank nodes, numeric literals, lists or quoted formulae.

\item{} A triple may be terminated by a period.
\begin{verbatim}
ExConf conf:eventName "WWW2006 Workshop on Models of Trust for the Web" .
\end{verbatim}

\item{} The repetition of another object for the same subject and
  predicate is possible using a comma ``,''
\begin{verbatim}
ExConf conf:cochair <http://csail.mit.edu/~lkagal/foaf#me>,
                        <http://umbc.edu/~finin/foaf#tim>.      
\end{verbatim}

\item{} The repetition of another predicate for the same subject is
  done using a semicolon ``;''
\begin{verbatim}
ExConf conf:eventName "WWW2006 Workshop on Models of Trust for the Web",
           conf:acronym "MTW06" .
\end{verbatim}

\item{} Blank nodes with certain properties can be defined by just
  putting the properties between ``[`` and ``]''. They can be used in two
  ways: [ ] together followed by the properties or [ ] around the
  properties. The following example describes the same subject in
  these two ways.

\begin{verbatim}
[ ] conf:homepage <http://www.l3s.de/~olmedilla/events/MTW06_Workshop.html> .
[ conf:homepage <http://www.l3s.de/~olmedilla/events/MTW06_Workshop.html> ] .
\end{verbatim}

\item{} N3 allows has a special \_: namespace prefix. An identifier of
  such a form (e.g. \_:xyz) represents an blank node.

\item{} An empty URI reference, which refers to the document it is
  written in, can be written using ``$<$$>$''. The example document
  declares itself to be a Conference as defined in the conf namespace.

\begin{verbatim}
<> a conf:Conference .
\end{verbatim}

\item{} Quoted formulae allow N3 formulae to be quoted within N3
  formulae using ``\{`` and ``\}''.  In a quoted N3 formula, s p o, both s
  and o can be RDF graphs or N3 formulae.

\begin{verbatim}
j:Joe says 
  { ex:ExConf conf:homepage <http://www.l3s.de/~olmedilla/events/MTW06_Workshop.html;
  conf:eventName "WWW2006 Workshop on Models of Trust for the Web";
  conf:acronym "MTW06" } .

\end{verbatim}

\item{} N3 formula can have both existential and universal
  quantifiers.  Existential variables can be indicated by an {\it
    @forSome} declaration and universal variables can be indicated by
  {\it @forAll} declarations. ?X is a shorthand notation and implies
  universal quantification in the enclosing parent of the current
  formula. It can be used without an explicit {\it @forAll}
  declaration. The scope of the @forAll, which is used to define
  universal variables, is outside the scope of any @forSome, which is
  used to define existential variables. If both universal and
  existential quantification are specified for the same context, then
  the scope of the universal quantification is outside the scope of
  the existentials.

\item{} Keywords are a very limited set of alphanumeric strings which
  are in the grammar prefixed by an ``@'' sign. If no {\it @keywords}
  directive is given, all qualified names need a ``:'' before them and
  all keywords except ``a'', ``is'', and ``of'' require an ``@''. If the
  @keywords directive is given, then the given set of bare strings
  (without either ``:'' or ``@'' before them) are keywords and the others
  are qualified names in the default namespace.

\item{} The keywords ``a'', ``is'', and ``of'' can be used without an ``@''
  even if the @keywords directive is not given. The keyword ``a'' maps
  to rdf:type whereas ``of'' and ``is'' provide syntactic sugar to
  describe triples in the reverse direction such as \{ {\it object} {\tt is}
  {\it predicate} {\tt of} {\it subject}\}.

\item{} The keywords {\it @true} and {\it @false} are boolean
  literals.

 \item{} Strings are defined within double quotes `` `` such as ``Joe
    Smith'' and within triple double quotes `` `` `` for multi-line values
    or values containing quote marks.

\item{} Numerical literals such as integers, floats, and decimals are
  also supported.

\item{} Comments are identified with ``\#''. Everything that follows the
  ``\#'' is ignored till the end of the line.

\item{} The shorthand =$>$ may be used for the {\it implies} property
  defined in the log: namespace
  (\url{http://www.w3.org/2000/10/swap/log#}). (Please refer to
  Section \ref{rules} for more information).

\item{} ``='' is a shorthand notation for the {\it sameAs} property
  defined in owl: namespace
  (\url{http://www.w3.org/2002/07/owl#}). (Please refer to Section
  \ref{relation} for more information).

\item{} N3 supports RDF collections and uses them frequently as
  ordered containers, as argument lists to N-ary functions such as
  {\it crypto:sign} in which the subject is a list of two things, a
  hash string and a key (containing private and public parts) and the
  object is calculated as a signature string by signing the hash with
  the key's private part.

For example, to describe Joe's interests we would use a list as
follows

\begin{verbatim}
j:Joe interests ( "AI" "Semantic Web" "Logic" ) .
\end{verbatim}

\end{itemize}

\comment{
\begin{itemize}

\item{} An {\it RDF triple} (or triple for short), described in N3 as
  s p o. , implies that some relationship defined by predicate, p,
  holds between the subject, s, and object, o. An example triple is

\begin{verbatim}
  <http://example.org/joe-foaf#joe>
  <http://xmlns.com/foaf/0.1/#schoolHomepage> <http://example.edu/> .
\end{verbatim}

\item{} A {\it namespace} is a collection of names, syntactically
  URIs, which share a common prefix. URIs are used as subjects,
  predicates, or objects in triples.

\item{} {\it Qualified names} or qnames are abbreviated names that are
  defined using declared namespaces such as {\it j:Joe} where is the
  namespace prefix for {\it http://example.org/joe-foaf\#}

\item{} An {\it RDF graph} is a set of triples and its meaning is
  defined as the logical conjunction of the triples it contains.
\begin{verbatim}
j:Joe foaf:schoolHomepage <http://example.edu> .
j:Joe foaf:currentProject proj:SemanticWebCalendar .
\end{verbatim}

\item{} The {\it RDF vocabulary} is a set of URI references in the
  rdf: namespace (\url{http://www.w3.org/1999/02/22-rdf-syntax-ns#})
  and defines a set of properties.

\begin{table} [h]
\begin{tabular}{p{12cm}}
\hline 

rdf:type rdf:Property rdf:XMLLiteral rdf:nil rdf:List rdf:Statement
rdf:subject rdf:predicate rdf:object rdf:first rdf:rest rdf:Seq
rdf:Bag rdf:Alt rdf:\_1 rdf:\_2 ... rdf:value \\

\hline
\end{tabular}
\caption{RDF vocabulary}
\label{rdfV-table}
\end{table}

\item{} An RDF {\it ontology} is a description of a world using the
  RDF vocabulary such as the FOAF ontology \cite{foaf}.

\item{} A {\it blank node} in an RDF graph indicates the existence of
  a thing with respect to its properties without actually naming or
  identifying it. It is considered an existential variable.

\item{} The {\it RDFS vocabulary} is a set of URI references in the
  rdfs: namespace (\url{http://www.w3.org/2000/01/rdf-schema#}) and
  defines classes.
 
\begin{table} [h]
\begin{tabular}{p{12cm}}
\hline 

rdfs:domain rdfs:range rdfs:Resource rdfs:Literal rdfs:Datatype
rdfs:Class rdfs:subClassOf rdfs:subPropertyOf rdfs:member
rdfs:Container rdfs:ContainerMembershipProperty rdfs:comment
rdfs:seeAlso rdfs:isDefinedBy rdfs:label\\

\hline
\end{tabular}
\caption{RDFS vocabulary}
\label{rdfsV-table}
\end{table}

\item{} In languages such as RDF/XML \cite{w3c04rdfxml} and
  OWL \cite{owl}, an empty URI reference always refers to the document
  that it is written in.

\end{itemize}
}

\end{document}